\documentclass[fleqn,usenatbib]{mnras}

\usepackage{newtxtext,newtxmath}

\usepackage[T1]{fontenc}

\DeclareRobustCommand{\VAN}[3]{#2}
\let\VANthebibliography\thebibliography
\def\thebibliography{\DeclareRobustCommand{\VAN}[3]{##3}\VANthebibliography}

\usepackage{graphicx}	
\usepackage{amsmath}	

\title[Alfvén wave modelling in a pseudo-streamer]{Density fluctuations in pseudo-streamer and helmet streamer: Generation, propagation and comparison with Metis observations}

\author[A. Kumar et al.]
{
A. Kumar,$^{1}$\thanks{E-mail: ak394@st-andrews.ac.uk}
S. Daley-Yates,$^{1}$
T. A. Howson,$^{2, 1}$
P. Pagano,$^{3, 4, 1}$
I. De Moortel,$^{1,5}$
and V. Andretta $^{6}$
\\
$^{1}$School of Mathematics and Statistics, University of St Andrews, St Andrews, Fife KY16 9SS, UK\\
$^{2}$Faculty of Design, Informatics and Business,
Abertay University, Bell Street, Dundee, DD1 1HG, UK
\\
$^{3}$Dipartimento di Fisica \& Chimica, Università degli Studi di Palermo, Via Archirafi 36, I-90123 Palermo, Italy\\
$^{4}$INAF-Osservatorio Astronomico di Palermo, Piazza del Parlamento 1, I-90134 Palermo, Italy\\
$^{5}$ Rosseland Centre for Solar Physics, University of Oslo, PO Box 1029 Blindern, NO-0315 Oslo, Norway\\
$^{6}$INAF–Osservatorio Astronomico di Capodimonte, Salita Moiariello 16, I-80131 Naples, Italy
}

\date{Accepted XXX. Received YYY; in original form ZZZ}

\pubyear{\the\year{}}

\begin{document}
\label{firstpage}
\pagerange{\pageref{firstpage}--\pageref{lastpage}}
\maketitle

\begin{abstract}
Recent white-light observations from the Metis coronagraph on board Solar Orbiter have shown the presence of propagating density fluctuations with a period of $\sim5$ minutes, particularly in the observationally bright regions of helmet streamers and pseudo-streamers. These density fluctuations have been shown to originate from magnetohydrodynamic wave modes -- slow and fast modes can directly compress the plasma, whereas Alfvén waves generate density fluctuations via a non-linear coupling with the slow and fast modes. In general, in low plasma $\beta$ regions, slow modes contribute more towards the generation of density fluctuations, whereas in high $\beta$ regions, there are significant contributions from both slow and fast waves. In this work, we investigate the propagation of (non-linear) Alfvén waves and the associated density fluctuations using a 2.5D axisymmetric global solar wind model that includes two distinct magnetic field topologies -- a helmet streamer and a pseudo-streamer. Key differences in these magnetic configurations give rise to differences in the wave behaviour due to, for example, the presence of a current sheet in the helmet streamer and an X-type null point in the pseudo-streamer, and the structure of the $\beta = 1$ layer in both magnetic field configurations. This leads to the possible creation of prominent interference patterns along the helmet streamer axis, which do not appear along the pseudo-streamer axis. Unlike in the streamer region, we also find that along the pseudo-streamer axis, the contribution to the density fluctuations is due to the fast waves only, despite being in a low $\beta$ plasma.
\end{abstract}


\section{Introduction}

The solar atmosphere is filled with magnetohydrodynamic (MHD) waves, which are expected to play an important role in the transport of energy. While there are three decoupled fundamental MHD wave modes in a homogeneous medium -- Alfvén, and slow and fast magnetoacoustic modes -- a direct observation of the fundamental wave modes is less likely due to the complex plasma distribution in the solar atmosphere. Instead, a combination of the fundamental modes or effects due to them are routinely observed \cite[e.g.][and references therein]{cirtain2007evidence, de2007chromospheric, banerjee2009signatures, banerjee2021magnetohydrodynamic, jess2023waves}. Furthermore, the presence of periodic (or quasi-periodic) fluctuations/inhomogeneities in the solar atmosphere suggests their wave origin. Some examples of these fluctuations/inhomogeneities are provided below.

While Alfvén waves are difficult to observe directly as they require simultaneous measurements of velocity and magnetic field fluctuations, \cite{kumar2026streamer} recently showed that through non-linear effects, they could be associated with density perturbations, observed as white-light intensity fluctuations. For example, recent white-light observations by the Metis coronagraph on board Solar Orbiter have revealed the presence of propagating 5-minute oscillations in helmet streamers as well as pseudo-streamers, extending as far as up to $\sim 2.3\, R_\odot$ in the middle corona \citep[][]{andretta2025first}. In this study, the authors detected fluctuations of the order of $\sim 0.1\%$ of the background density in the corona, propagating with speeds ranging in $150-450\, \mathrm{km\, s^{-1}}$ measured along different paths in the Metis field of view. 

The 5-minute periodicity ($3\, \mathrm{mHz}$ frequency) associated with the density fluctuations in \cite{andretta2025first} corresponds to the dominant periodicity of the global internal pressure waves (\textit{p-}modes) of the Sun. Hence, these authors suggest that the observed density fluctuations are due to the \textit{p-}modes. On the global scale, gravitational stratification imposes an acoustic cutoff such that frequencies lower than $3\, \mathrm{mHz}$ are filtered out and only higher frequencies can propagate through the upper layers of the solar atmosphere \citep[e.g.][and references therein]{felipe2020numerical}. However, the acoustic cutoff can vary if the magnetic field strength and its inclination with respect to the direction of wave propagation are changed \citep[][]{jefferies2006magnetoacoustic, stangalini2011mhd}. As the purely acoustic \textit{p-}mode waves reach upper layers of the atmosphere, they can mode convert into Alfvén and slow magnetoacoustic wave modes \citep[e.g.][and references therein]{schunker2006magnetic, cally2008three, khomenko2012numerical} that can produce wave-like fluctuations with a corresponding period.

Apart from \cite{andretta2025first}, observations of transverse coronal oscillations at a similar frequency range have been discussed by several authors. See for example, \cite{morton2016global}, who use measurements from the Coronal Multi-Channel Polarimeter (CoMP), in the corona below $1.3\, R_\odot$. While Alfvén waves remain challenging to detect from remote observations, previous studies have linked such oscillations with the injection of Alfvénic waves in the corona \citep[][]{morton2019basal}. Recent observations from the Cryo-NIRSP on board the Daniel K. Inouye Solar Telescope (DKIST) have also shown the presence of high frequency ($> 10\, \mathrm{mHz}$) Alfvénic waves as well as the presence of small amplitude torsional Alfvén waves \citep[][]{morton2025evidence, morton2025high}. In the solar wind, Parker Solar Probe (PSP) observations of Alfvénic wave trains with periods around $5$ minutes at $\sim 9.9\, R_\odot$ have also been reported \citep[][]{huang2025situ}. Furthermore, \textit{p-}modes can also modulate magnetic reconnection to generate short period periodic density structures in the heliosphere \citep[e.g.][]{viall2008inherent, kepko2024periodic}.

\cite{kumar2026streamer} considered the role of different MHD wave modes -- Alfvén and slow and fast magnetoacoustic -- to generate 5-minute propagating density fluctuations in a helmet streamer, as in \cite{andretta2025first}. Density fluctuations can be generated directly from the slow and fast magnetoacoustic modes, or indirectly from Alfvén waves undergoing mode conversion to the compressive slow and fast modes. Alfvén waves can mode convert via non-linear ponderomotive effects \citep[][]{hollweg1982alfven, verwichte1999evolution}, non-linear phase mixing \citep[][]{heyvaerts1983coronal, nakariakov1997alfven, botha2000developed, thurgood2013ponderomotive}, or a combination of both. Using both Alfvén and magnetoacoustic waves in their simulations, \cite{kumar2026streamer} demonstrated that the dynamical behaviour of the resulting density fluctuations is similar in both cases, with slow modes predominantly contributing to the generation of density fluctuations in a low $\beta$ regime, and significant contributions from both slow and fast modes in the high $\beta$ regime. Density fluctuations in the solar wind are also known to be associated with the formation of solar wind turbulence. Notably, combined PSP and Solar Orbiter observations find evidence of transonic turbulence associated with the observed density fluctuations in the near-Sun solar wind (between approximately $11$ and $80\, R_\odot$), which \cite{zhao2025transonic} suggest could be associated with fast magnetoacoustic waves.

In this paper, we study the propagation of non-linear Alfvén waves and associated density fluctuations in two distinct magnetic topologies -- a helmet streamer and a pseudo-streamer -- in the same simulation. A helmet streamer consists of a single system of coronal loops that extend outwards to form a current sheet that separates opposite polarity magnetic field lines. In contrast, a pseudo-streamer consists of two adjacent coronal arcades that open into the same magnetic polarity, with the open field lines separated by a separatrix. As such, there is no current sheet associated with a pseudo-streamer, but there is instead an X-type null point (in two-dimensions) above the closed loops. 

By having such a combined setup, we extend our previously reported model in \cite{kumar2026streamer} to include more complex and realistic magnetic field geometries. This enables us to not only directly compare our simulations with Metis observations \citep[][]{andretta2025first}, but also to explore the propagation of Alfvén waves and density fluctuations in these different magnetic topologies. The paper is structured as follows. In Section \ref{sec_setup}, we describe the numerical model, the initial and boundary conditions and the Alfvén wave driver. The results are presented in Section \ref{sec_results}, with a separate discussion on the propagation of waves and density fluctuations in a helmet streamer (Section \ref{sec_hs}), a pseudo-streamer (Section \ref{sec_ps}) and their observational signatures (Section \ref{sec_slits}). Finally, discussion and conclusions are provided in Section \ref{sec_discussion}.

\section{Numerical setup} \label{sec_setup}
The numerical simulation presented in this paper investigates the propagation of Alfvén waves in a steady background containing both a helmet streamer and a pseudo-streamer. This 2.5 dimensional model builds on the model presented in \cite{kumar2026streamer}, which only included a helmet streamer. The simulation is set up to describe the solar wind in spherical geometry within an MHD framework under the effect of gravity. The energy transport is further governed by thermal conduction, optically-thin radiative cooling, uniform resistivity ($\eta = 10^{15}\, \mathrm{cm^2\, s^{-1}}$) and a background heating function that supports coronal temperatures and a super-Alfvénic solar wind. As such, the following MHD equations are advanced in PLUTO \citep[][]{mignone2007pluto}:
\begin{gather}
    \frac{\partial \rho}{\partial t} + \mathbf{\nabla}\cdot \left(\rho \mathbf{v}\right) = 0, \label{eq_continuity} \\
    \rho \frac{d \mathbf{v}}{dt} = -\mathbf{\nabla}p + \mathbf{j}\times \mathbf{B} + \rho \mathbf{g} + \rho \mathbf{S}_\mathrm{w} \left(r, \theta, t\right), \label{eq_momentum} \\
    \frac{\partial \mathbf{B}}{\partial t} = \mathbf{\nabla}\times \left(\mathbf{v} \times \mathbf{B}\right) + \eta \nabla^2 \mathbf{B}, \label{eq_induction} \\
    \frac{\rho ^\gamma}{\gamma - 1}\frac{d}{dt}\left(\frac{p}{\rho^\gamma} \right) = -\mathbf{\nabla}\cdot \mathbf{q} - n^2 \Lambda\left( T \right) + \eta j^2 + Q_{\mathrm{bg}} \label{eq_energy}, \\
    p = \frac{\rho k_B T}{\mu m_p}. \label{eq_ideal_gas}
\end{gather}
Here, $\rho$, $p$, $\mathbf{v}$ and $\mathbf{B}$ denote the density, gas pressure, bulk velocity and the magnetic field respectively and the current density is $\mathbf{j}=\mathbf{\nabla}\times \mathbf{B}/\mu_0$, with $\mu_0$ being the magnetic permeability of free space. The ideal gas law (\ref{eq_ideal_gas}) provides closure to the system of equations, where $k_B$, $T$, $m_p$ and $\mu$ denote Boltzmann's constant, temperature, proton mass and the mean molecular weight (chosen to be $0.63m_p$). The above system of equations have two ad hoc terms -- $\mathbf{S}_\mathrm{w}$ and $Q_\mathrm{bg}$ -- that denote a wave source term and the background heating function; the former is described in detail in Section \ref{sec_wave_driver}. The latter, is the heating function in (\ref{eq_energy}) and is defined as in \cite{reville2018parametric}:
\begin{equation}
    Q_\mathrm{bg} \left(r\right) = \frac{f_0}{H_0} \left(\frac{R_\odot}{r}\right)^2 \exp \left(- \frac{r - R_\odot}{H_0} \right), \label{eq_heating}
\end{equation}
where the scale height ($H_0$) and the photospheric energy flux ($f_0$) are $1\,R_\odot$ and $1.5\times 10^5\, \mathrm{erg\,cm^{-2}\,s^{-1}}$ respectively, and $r$ denotes the radial coordinate. A strong chromospheric heating applied via this function generates large upflows, which deposits plasma higher up in the domain, eventually forming the solar wind.

In this model, thermal conduction is modelled using the usual Spitzer-Härm approach where the conductive flux is split into parallel and perpendicular components (with respect to the magnetic field):
\begin{equation}
    \mathbf{\nabla}\cdot\mathbf{q} =
\kappa_{\parallel} \hat{\mathbf{B}} (\hat{\mathbf{B}}\cdot\nabla T) + \kappa_{\perp} \left[ \nabla T - \hat{\mathbf{B}} (\hat{\mathbf{B}} \cdot \nabla T) \right].
\end{equation}
In our model, we have used $\kappa_\parallel = 9\times10^{-7} T^{5/2}\, \mathrm{erg\, cm^{-1} \, s^{-1}\, K^{-1}}$ and $\kappa_\perp = 0$, since heat conduction in a magnetised plasma is highly anisotropic \citep[][]{priest2014magnetohydrodynamics}. The radiative cooling function, $\Lambda\left(T\right)$, is derived from \cite{klimchuk2008highly}.

Since the physical domain of our simulation includes a $5\, \mathrm{Mm}$ wide chromospheric layer followed by a narrow transition region, we modify the conduction tensor and the cooling function as follows:
\begin{gather}
    \kappa_\parallel = \begin{cases}\kappa_0 T^{5/2},\ T\geq T_M,\\
    \kappa_0 T_M^{5/2},\ T< T_M
                        \end{cases}, \\
    \Lambda\left(T\right) = \begin{cases} \Lambda\left(T\right),\ T\geq T_M,\\
    \Lambda\left(T\right)\left(\frac{T}{T_M}\right)^{5/2},\ T< T_M
                            \end{cases},
\end{gather}
with $T_M = 2\times 10^5\, \mathrm{K}$, following \cite{lionello2008multispectral} and \cite{mikic2013importance}. This ensures a broader transition region, allowing us to resolve the energy flux through this region. 

\subsection{Initial conditions}
\label{sec_initial_conditions}
The physical domain of our simulation starts from the chromosphere at $1\, R_\odot$ and extends up to $20\, R_\odot$. The isothermal and hydrostatic chromosphere is initialised with a temperature of $10^4\, \mathrm{K}$ and a number density of $10^{12}\, \mathrm{cm^{-3}}$, which drops to $\sim 10^9\, \mathrm{cm^{-3}}$ at $5\, \mathrm{Mm}$ above the surface. This base density is used together with a coronal temperature of $1.7\times10^{6} \ \mathrm{K}$ to initialise a Parker isothermal wind solution above the chromospheric layer \citep[][]{parker1958dynamics}.

The initial magnetic field configuration consists of a localised bipolar flux patch -- located at $60^\circ$ from the north pole -- embedded in a global dipole ($\mathbf{B}_\mathrm{global}$):
\begin{gather}
    \mathrm{B}_0\left(r, \theta\right) = \mathbf{B}_\mathrm{global}+ \mathbf{B}_\mathrm{bip} = \frac{2b_0\cos\theta}{r^3}\, \hat{\mathbf{r}} + \frac{b_0\sin\theta}{r^3}\, \hat{\mathbf{\theta}} + \mathbf{B}_\mathrm{bip}. \label{eq_mag_field}
\end{gather}
In order to construct this bipolar flux patch ($\mathbf{B}_\mathrm{bip}$), we assume that it is initially potential, meaning that it can be expressed in terms of a scalar potential ($\Psi_\mathrm{bip}$) as follows:
\begin{gather}
    \mathbf{B}_\mathrm{bip} \left(r, \theta \right) = \mathbf{\nabla}\Psi_\mathrm{bip} \left(r, \theta \right) \label{eq_scalar_potential}.
\end{gather}
In spherical geometry with axial symmetry, the scalar potential is expressed as a linear combination of mutually orthogonal associated Legendre polynomials ($P_l$) as
\begin{gather}
    \Psi_\mathrm{bip} \left(r, \theta\right) = \sum_{l = 0}^\infty \frac{b_l}{r^{l+1}}P_l\left(\cos \theta\right), \label{eq_psi_legendre}
\end{gather}
where the subscript $l$ denotes the $l$th Legendre polynomial, weighted by the coefficient $b_l$ \citep[][]{priest2014magnetohydrodynamics}; since we are seeking a magnetic field configuration with a radially decaying strength, the coefficients that correspond to a radially growing field strength are omitted in (\ref{eq_psi_legendre}). The coefficients $b_l$ are obtained by imposing appropriate boundary conditions, which, in our case, would be a flux distribution at $r = 1\, R_\odot$. The chosen flux distribution consists of two regions with opposite polarities, with polarity inversion at $60^\circ$ and a total width of $10^\circ$, i.e., the radial magnetic field at $r = 1\, R_\odot$ is expressed by
\begin{gather}
    B_{r_\mathrm{bip}}\left(r = 1\, R_\odot, \theta\right) = \tanh\left(\pi\left(\frac{\theta - 60^\circ}{180^\circ}\right) \right)\exp\left(-\left(\frac{\theta - 60^\circ}{10^\circ} \right)^2 \right). \label{eq_surface_flux_bip}
\end{gather}

\begin{figure}
    \centering
    \includegraphics[width=\linewidth]{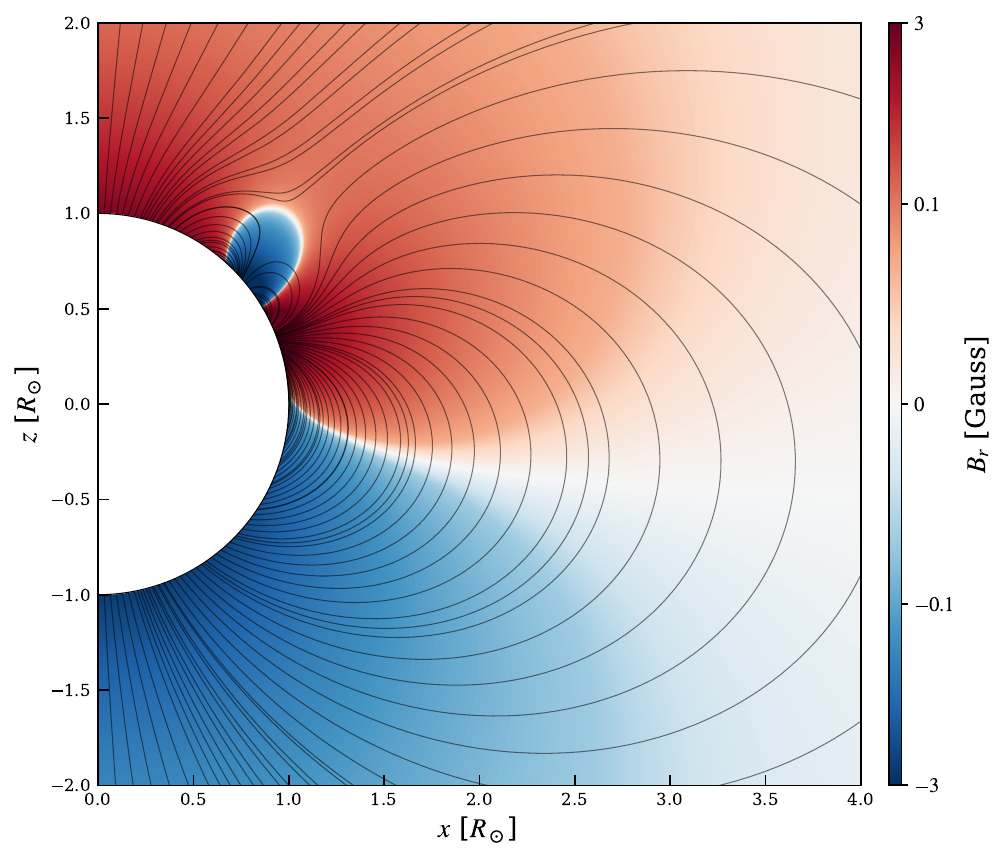}
    \caption{Initial magnetic field configuration used for a combined simulation of a helmet streamer and a pseudo-streamer. The colours denote the radial magnetic field ($B_r$), whereas the solid black lines denote the magnetic field lines. The initial magnetic field comprises of a small bipolar flux patch embedded in the northern hemisphere of the global dipole; the former extends into a pseudo-streamer, while the latter forms the helmet streamer.}
    \label{fig_initial_field}
\end{figure}

This radial magnetic field at $r = 1\, R_\odot$ (\ref{eq_surface_flux_bip}) is matched with the field obtained by substituting (\ref{eq_psi_legendre}) in (\ref{eq_scalar_potential}) to yield the coefficients $b_l$. Subsequently, the scalar potential is reconstructed and the magnetic field profile in the entire domain is obtained by taking its gradient (\ref{eq_scalar_potential}). We set the dipole strength to be $1\, \mathrm{Gauss}$ at the equator and scale the embedded bipole so that it has a maximum strength of $3\, \mathrm{Gauss}$. Since a pseudo-streamer consists of two coronal arcades with the same magnetic field polarity in the boundaries, the orientation of the imposed bipolar flux patch is opposite to that of the global dipole. The initial magnetic field configuration is shown in Fig. \ref{fig_initial_field}.

\subsection{Boundary conditions and numerical grid}
At the lower radial boundary, we fix the density, pressure magnetic field and velocity in the ghost cells. In a layer of cells (equal to the number of ghost cells) above the boundary, we modify the velocity such that it is parallel to the magnetic field in those cells. Lastly, $\mathrm{d}B_\phi/\mathrm{d}r = \mathrm{constant}$ sets the condition on $B_\phi$. At the outer radial boundary, $\rho r^2$, $p r^2$, $\rho v_r$, $v_\theta$, $rv_\phi$, $B_r r^2 $, $B_\theta$ and $rB_\phi$ are all held constant. The boundary conditions in the $\theta-$direction are set to PLUTO's predefined `axisymmetric' boundaries.

Our simulation is advanced on a two-dimensional grid in spherical co-ordinates with meridional symmetry on the $r-\theta$ plane (i.e. the simulation is axisymmetric in the $\phi-$direction). The radial grid starts at $1\, R_\odot$ and extends until $20\, R_\odot$, consisting of $2816$ cells. The chromosphere and the transition, up to $10\, \mathrm{Mm}$, are resolved using $128$ cells that are uniformly spread. This is followed by a stretched grid (where the stretching factor is calculated using the adjacent uniform grid to ensure a smooth transition) until $2.4\, R_\odot$, which uses $1536$ cells, whereas the remaining $1152$ cells are logarithmically stretched until $20\, R_\odot$. In the $\theta-$direction, we have used $1920$ cells that are uniformly space between $0^\circ$ and $180^\circ$.

The time evolution uses the second order Runge-Kutta scheme, and we have used linear spatial reconstruction via the van Leer slope limiter and the Harten, Lax, van Leer (HLL) Riemann solver. Finally, the divergence-free condition is imposed using Powell's eight-wave formalism, and resistivity and thermal conduction are handled using the second order RK-Legendre scheme.

\subsection{Relaxation and evolution to a combined pseudo-streamer - helmet streamer configuration}
\begin{figure*}
    \centering
    \includegraphics[width=\linewidth]{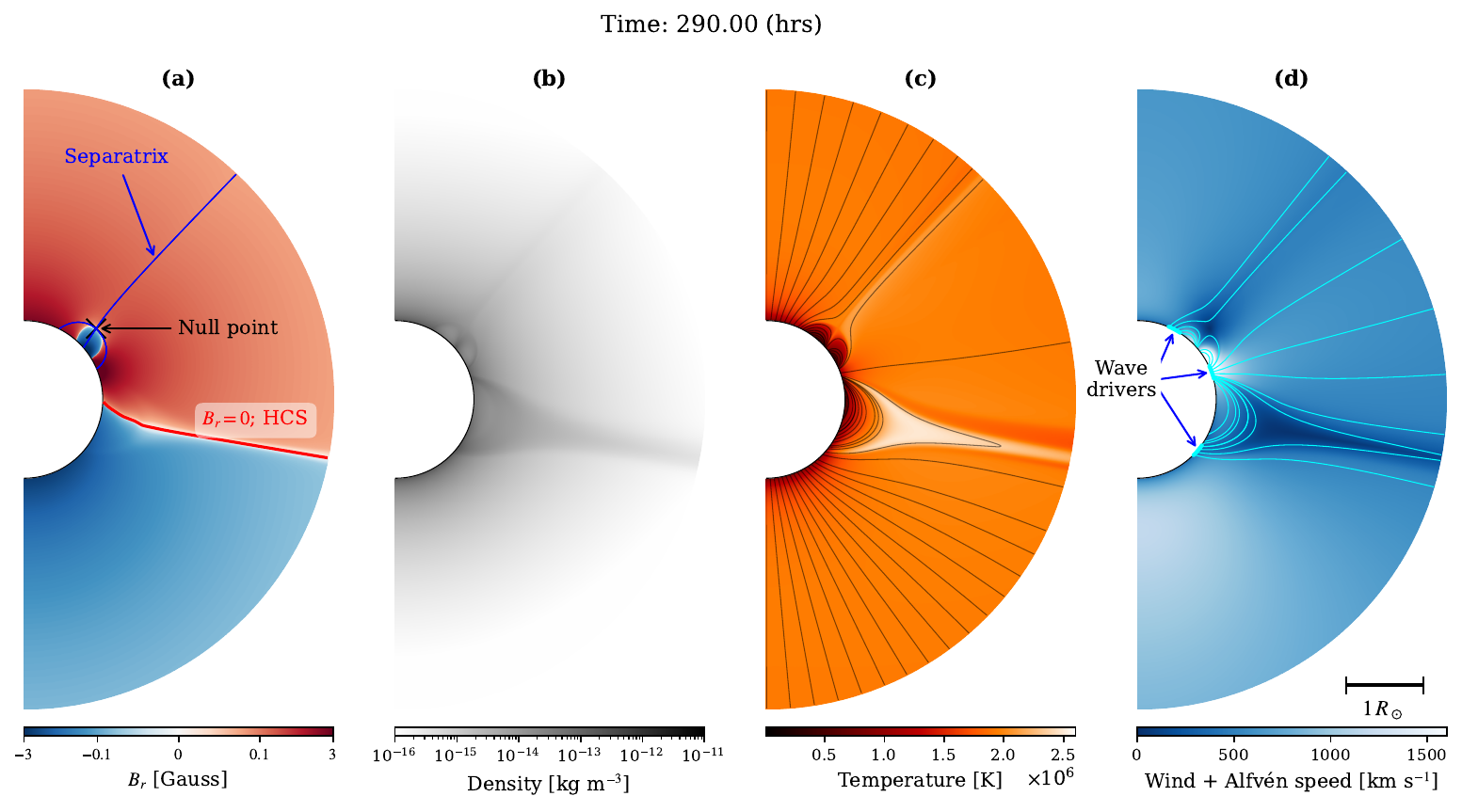}
    \caption{Steady-state plasma configuration (attained after $\sim 12$ days of simulated time) used for further wave simulations, with the colour maps denoting \textbf{(a)} the radial magnetic field ($B_r$), \textbf{(b)} density ($\rho$), \textbf{(c)} temperature ($T$), and \textbf{(d)} Alfvén wave propagation speed -- sum of the Alfvén speed ($v_A$) and the solar wind speed ($v_\mathrm{sw}$). The solid red line in (a) denote the $B_r = 0$ region associated with the helmet streamer, forming the heliospheric current sheet (HCS). On the other hand, the null point and the separatrix associated with the pseudo-streamer are shown by the black cross and the solid blue line respectively. The solid black lines in (c) denote the magnetic field lines. Finally, the blue patches in (d) denote the locations of Alfvén wave drivers, with the magnetic field lines traced from those regions shown in by the same colour.}
    \label{fig_steady_state}
\end{figure*}

The initial conditions of Parker's solar wind are not in equilibrium with the imposed magnetic field. Therefore, we evolve this setup numerically in time, allowing it to relax and settle into a steady state before waves are injected. The imposed background heating leads to plasma evaporation from the chromosphere, eventually forming a fast, hot and dense solar wind. This plasma outflow stretches closed magnetic field lines near the equator, which eventually form a helmet streamer, and in the northern hemisphere, where a pseudo-streamer forms. In the closed field regions of the helmet streamer and pseudo-streamer, we see the build up of hot and dense plasma, where both the temperature and the density are higher in the helmet streamer than in the pseudo-streamer. In the open field regions -- near the poles and in between the helmet streamer and the pseudo-streamer -- the plasma flows along the near-radial magnetic field.

\begin{figure}
    \centering
    \includegraphics[width=\linewidth]{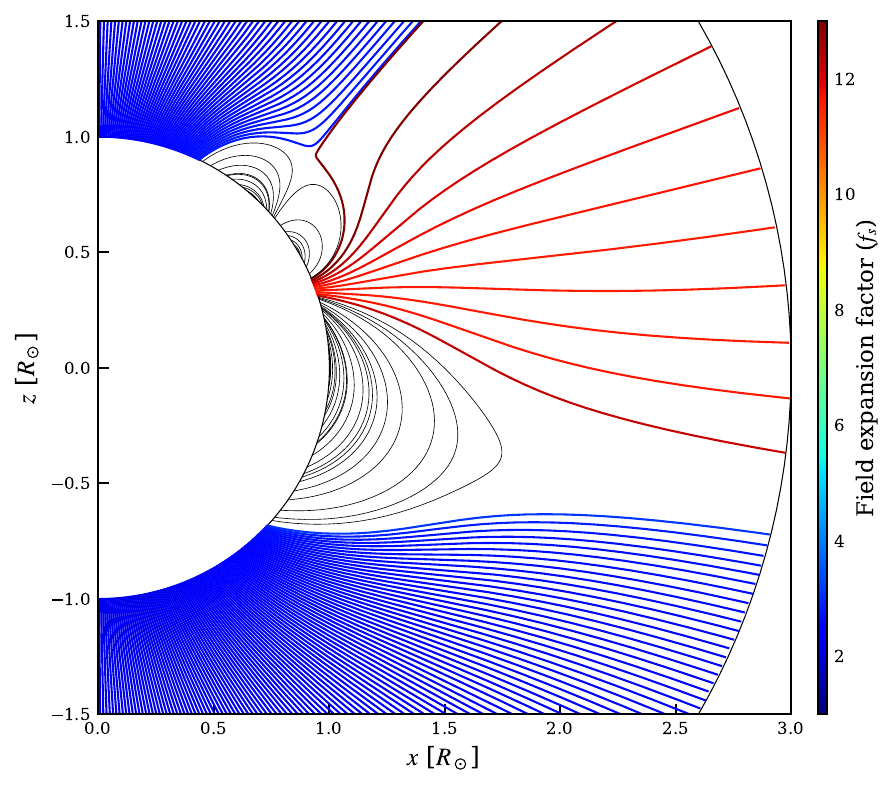}
    \caption{Expansion factor of the open magnetic field lines in steady-state. The open field lines are coloured according to the expansion factor, calculated using (\ref{eq_expansion_factor}). Here, we have considered the inner boundary at $r_1 = 1\, R_\odot$ and the source surface at $r_2 = 3\, R_\odot$, where the field lines are truncated. The closed lines, in black, are shown here for completeness.}
    \label{fig_expansion_factor}
\end{figure}

The simulation attains a steady state after $\sim 12$ days of simulated time, shown in Fig. \ref{fig_steady_state}. This steady state exhibits the combined helmet streamer and a pseudo-streamer topology described in Section \ref{sec_initial_conditions}. In the steady state, the wind speed in the open field regions at $\sim 4\, R_\odot$ reaches $\sim 200\, \mathrm{km\, s^{-1}}$, whereas in the closed field regions of the helmet streamer and the pseudo-streamer, the wind speed is $\sim 50\, \mathrm{km\, s^{-1}}$. We note that the resultant steady-state solar wind speeds are dependent on the heating function (i.e., the values of $H_0$ and $f_0$ in (\ref{eq_heating})) as well as the initial magnetic field configuration, which determine the plasma distribution in the simulation domain. Therefore, the characteristic phase speeds -- Alfvén, slow and fast magnetoacoustic speeds -- are also dependent on these factors. The values of $H_0$ and $f_0$ are taken from the literature \citep[][]{reville2018parametric}, to give realistic values for the solar wind speed and density in our simulation domain.

The presence of the pseudo-streamer tilts the helmet streamer south of the equator, resulting in a region of high magnetic field expansion between the pseudo-streamer and the helmet streamer. This is shown in Fig. \ref{fig_expansion_factor}, where each field line is coloured according to the expansion factor ($f_s$), defined as:
\begin{equation}
    f_s = \left(\frac{r_1}{r_2}\right)^2 \frac{B_r(r_1)}{B_r(r_2)}, \label{eq_expansion_factor}
\end{equation}
where $r_1 = 1\, R_\odot$ and $r_2 = 3\, R_\odot$ are the chosen reference points where the radial magnetic field is measured \citep[][]{wang1990solar}; $f_s = 1$ for a radial expansion whereas $f_s > 1$ implies a super-radial expansion. In our simulation, the open magnetic field lines expand super-radially everywhere in the domain, with the maximum expansion in the region between the pseudo-streamer and the helmet streamer.

The helmet streamer has a single magnetic arcade with open field lines of opposite polarities on either side, thus forming a current sheet (marked by the red contour in Fig. \ref{fig_steady_state}(a)). In contrast, the pseudo-streamer has a double arcade bounded by open field lines of the same polarity on either side. Therefore, there is no current sheet associated with a pseudo-streamer, but there is an X-type null point near the apex of the closed field lines (marked by a black cross in Fig. \ref{fig_steady_state}(a)). The steady-state magnetic field configuration as well as some of the plasma quantities are shown in Fig. \ref{fig_steady_state}. This is the setup that we use for the wave simulation, described below.

\subsection{Wave drivers}\label{sec_wave_driver}
\begin{figure}
    \centering
    \includegraphics[width=\linewidth]{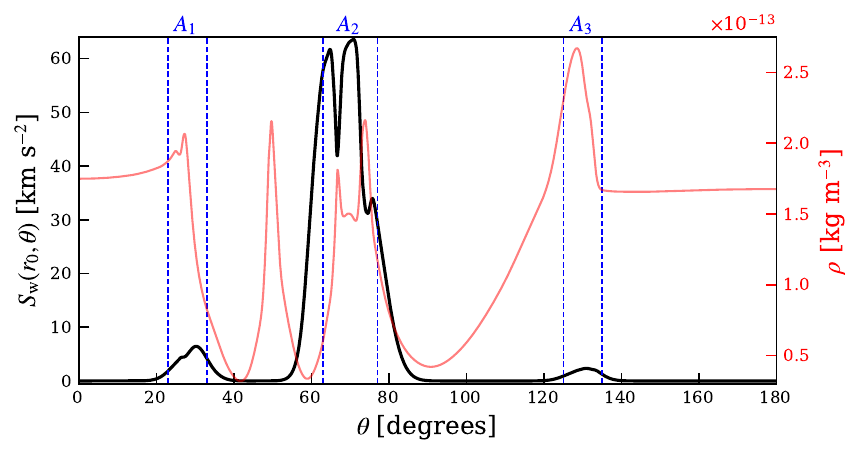}
    \caption{Acceleration profile for the Alfvén wave driver (shown in black) in the latitudinal ($\theta$) direction. The driver locations are marked by the dashed blue lines, covering footpoints of the helmet streamer as well as the pseudo-streamer, including the region between them. The solid red line denotes the density profile in the driver region, provided here for completeness, since the resulting Alfvén wave amplitude will also depend on the density due to this momentum driver.}
    \label{fig_driver_amplitude}
\end{figure}

In this subsection we describe the Alfvén wave driver source term that appears in (\ref{eq_momentum}), $\mathbf{S}_\mathrm{w}$. Since both (non-linear) Alfvén waves and magnetoacoustic waves produce qualitatively similar density fluctuations, as demonstrated in {\cite{kumar2026streamer}, we only consider Alfvén waves in this simulation, i.e., $\mathbf{S}_\mathrm{w} = S_\mathrm{s}\, \hat{\phi}$. Furthermore, the analysis of the parallel and perpendicular velocity perturbations, associated with the (non-linearly) generated slow and fast waves \citep[see][]{kumar2026streamer} is simpler as the driven Alfvén perturbations will be in the $\phi$ plane and hence easily distinguished from the slow and fast wave perturbations in the ($r-\theta$) plane. We simultaneously excite Alfvén waves at three distinct locations near the lower boundary: (i) north of the pseudo-steamer, (ii) in between the pseudo-streamer and the helmet streamer, and (iii) south of the helmet streamer (see Fig. \ref{fig_steady_state}(d)). This arrangement of drivers injects waves from both footpoints of both the streamer and the pseudo-streamer, capturing both open and closed field lines. The wave source term is expressed as:
\begin{gather}
    S_\mathrm{w} \left(r, \theta, t\right) = \sum_{i = 1}^{i = 3} A_i \exp\left[-\left(\frac{r - r_0}{\sigma_r} \right)^2 \right] \exp \left[ -\left(\frac{\theta - \theta_i}{\sigma_{\theta_i}} \right)^2 \right] \sin \omega t, \label{eq_driver}
\end{gather}
where $r_0 = 1.025\, R_\odot$ and $\sigma_r = 5\times 10^{-4}\, R_\odot$ define the centre and width of the wave driver in the radial direction. Similarly, $\theta_i$ and $\sigma_{\theta_i}$ define the driver's centre and width in the latitudinal direction, with $i$ labelling each driver. We used $\theta_i \in \{28^\circ, 70^\circ, 130^\circ \}$, $\sigma_{\theta_i} \in \{5^\circ, 7^\circ, 5^\circ\}$ for $i = 1, 2, 3$ respectively; the regions of Alfvén wave excitation are marked by the blue patches in Fig. \ref{fig_steady_state}(d) and the blue lines denote the magnetic field lines emanating from the driven region. The angular frequency ($\omega = 2\pi/P$) is set by a period $P$ of $10$ minutes which, due to the action of the ponderomotive force, generates periodic density fluctuations with half the period \citep[][]{kumar2026streamer}. Furthermore, the amplitudes $A_i$ have units of acceleration, therefore the resultant wave amplitude also depends on the density in the driver region. Since the density varies with latitude due to the presence of different magnetic structures, we tune each acceleration amplitude such that the resulting Alfvén waves have approximately equal amplitudes away from the driver region; these amplitudes are $\sim 30\, \mathrm{km\, s^{-1}}$, consistent with coronal Alfvénic fluctuations reported by e.g.~\cite{mcintosh2012estimating, weberg2018automated, morton2019basal}. The profile of the wave driver amplitude (and also of the density for reference) as a function of latitude is shown in Fig. \ref{fig_driver_amplitude}; note that the highest amplitude is for the driver between the pseudo-streamer and the helmet streamer, where the expansion factor ($f_s$) is also the highest (see Fig. \ref{fig_expansion_factor}).  

Alfvén waves are injected into the domain through the wave driver after a steady state has been reached. As such, the simulation is restarted and continued for $12$ hours of simulated time, with outputs every $30$ seconds, giving 20 snapshots per wave driver cycle. The timestamps shown in subsequent figures denote the time corresponding to this wave simulation (i.e. $t=0$ corresponds to the start of the wave driving). Additionally, since our region of interest lies in the middle corona, the figures are truncated to $\sim 4\,R_\odot$ in most cases, and projected onto Cartesian coordinates.

\section{Results} \label{sec_results}
\begin{figure*}
    \centering
    \includegraphics[width=\linewidth]{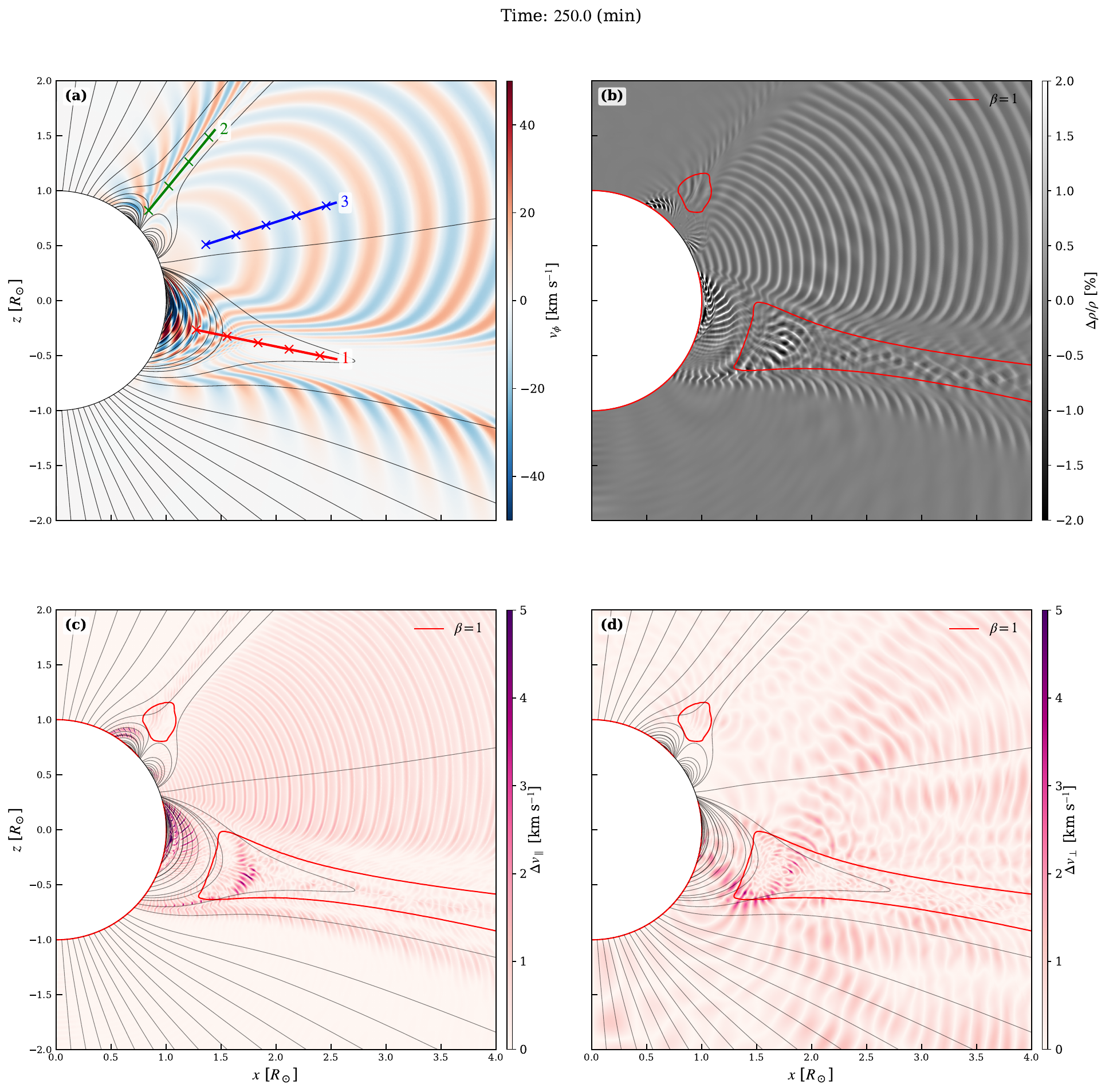}
    \caption{Wave simulation output at an intermediate time, showing the propagation of \textbf{(a)} Alfvén waves ($v_\phi$), \textbf{(b)} density fluctuations ($\Delta \rho/\rho$), and Alfvén wave-induced flows in the simulation plane \textbf{(c)} parallel ($\Delta v_\parallel$) and \textbf{(d)} perpendicular ($\Delta v_\perp$) to the magnetic field lines. The parallel and perpendicular velocities indicate the slow and fast magnetoacoustic modes, generated from the driven Alfvén waves. The solid black lines denote the magnetic field lines, and the red contours denote the $\beta = 1$ layer in the simulation domain; note that the high $\beta$ region for the helmet streamer stretches along its current sheet, whereas the high $\beta$ region for a pseudo-streamer is localised above the closed loops, embedding the null point. The straight lines numbered 1 to 3 in (a) denote the slits used for further analysis (Section \ref{sec_slits}) and the crosses are drawn at every $200\, \mathrm{Mm}$ for ease of comparison with subsequent time-distance plots.}
    \label{fig_alfven_rho}
\end{figure*}

Alfvén waves in an homogeneous medium in the linear regime are incompressible, and, in the linear regime, they are decoupled from the slow and fast magnetoacoustic modes. As soon as these idealised conditions of homogeneity and linearity are lifted, Alfvén waves can indirectly generate slow and fast magnetoacoustic wave modes through the ponderomotive or (non-linear) phase mixing. For example, an inhomogeneity in the direction transverse to the magnetic field predominantly generates fast modes via (non-linear) phase mixing \citep[e.g.][and references therein]{nakariakov1997alfven, thurgood2013ponderomotive, davies2025alfven}, and non-linear Alfvén waves can generate both slow and fast mode waves through the ponderomotive effect \citep[e.g.][]{verwichte1999evolution, kumar2026streamer}. In the simulation discussed in this article, density fluctuations are generated through a combination of these effects, since the plasma background used in our simulation is inhomogeneous and the driven Alfvén waves are in the non-linear regime. To determine whether these density perturbations are due to the slow or fast wave mode, we use the decomposition into parallel and perpendicular components of the associated velocity perturbations (see for example, \cite{kumar2026streamer} but also \citet{mclaughlin2006mhd}}.

In general, as Alfvén waves propagate through the domain, they encounter gradients, both along and across the magnetic field, in the propagation speed (due to variations in the local field strength, plasma density and the solar wind speed). For example, Alfvén waves undergo continuous reflections in regions where there are field-aligned changes in plasma density. Similarly, the generated density fluctuations undergo continuous reflections and refraction, leading to more complex wave dynamics \citep[][]{kumar2026streamer}. A snapshot of the simulation at $t = 250\, \mathrm{min}$ is provided in Fig. \ref{fig_alfven_rho}, with the four panels denoting (a) the Alfvén wave component, (b) the running difference of the density from two successive simulation outputs $30\, \mathrm{s}$ apart (c) parallel, and (d) perpendicular components (to the magnetic field, denoted by the solid black lines) of the induced velocities in the simulation plane due to the driven Alfvén waves; the parallel and perpendicular components are interpreted as slow and fast wave components respectively, as in \cite{mclaughlin2006mhd}.

A detailed discussion on the propagation of Alfvén waves and density fluctuations is provided in the following sections. In Sections \ref{sec_hs} and \ref{sec_ps}, we discuss the wave dynamics in the different magnetic field topologies of a helmet streamer and a pseudo-streamer, respectively. This is followed by a direct comparison with Metis observations of apparent propagation speeds along different slits in the simulation domain in Section \ref{sec_slits}.

\subsection{Wave propagation in the helmet streamer} \label{sec_hs}
Alfvén waves driven at the footpoints of the helmet streamer propagate into the domain as they are guided by the magnetic field; their propagation and the dynamics associated with them are characterised by the local plasma conditions. In the open field regions -- at the two flanks of the helmet streamer -- the Alfvén wave amplitude increases from $\sim 10 \, \mathrm{km\, s^{-1}}$ in the low corona, to $\sim 20\, \mathrm{km\, s^{-1}}$ at $\sim 2.5\, R_\odot$ (see Fig. \ref{fig_alfven_rho}(a)). This amplitude increase is due to a monotonic decrease in density -- due to gravitational stratification -- along those open field lines. On the other hand, along the closed loops, Alfvén waves reflect from the opposite footpoint to create interference patterns along the loop (Fig. \ref{fig_alfven_rho}(a)) with wave amplitudes reaching up to $\sim 50\, \mathrm{km\, s^{-1}}$ at $\sim 1.5\, R_\odot$. In the streamer cusp, there is a decay of the Alfvén wave amplitudes, primarily due to a steep decline in the magnetic field strength. Consequently, the Alfvén speed, and therefore the wavelengths, also reduce. This leads to some numerical dissipation as the waves become poorly resolved. This results in a loss of Alfvén wave energy in the majority of the current sheet, as waves from opposite footpoints remain confined to the open field lines, with interference only in the closed loops below the dense helmet streamer core.

As in \cite{kumar2026streamer}, the driven Alfvén waves generate slow and fast waves via the ponderomotive force and (non-linear) phase mixing (see Figs. \ref{fig_alfven_rho}(a) and (b), respectively). As they are compressive, the slow and fast wave modes are associated with both density and velocity perturbations. As the slow and fast wave characteristics change for different plasma $\beta$ regimes, it is helpful to decompose the perturbed velocities into components that are parallel ($\Delta v_\parallel$) and perpendicular ($\Delta v_\perp$) to the magnetic field, shown in Fig. \ref{fig_alfven_rho}(c) and (d) respectively. As discussed extensively in \cite{kumar2026streamer}, the contribution of slow and fast modes to the density fluctuations varies with the plasma $\beta$; in the low $\beta$ regime, there is a higher contribution from the slow modes, and hence, the propagation of density fluctuations is predominantly tied to the magnetic field lines. On the other hand, in a high $\beta$ regime, both slow and fast modes contribute to the density fluctuations, and propagation becomes isotropic \citep[see][Appendix B]{kumar2026streamer}. We see a similar effect in the simulation in this study, where in the low $\beta$ regime -- at the two flanks of the helmet streamer as well as below the helmet streamer cusp -- propagation of the density fluctuations is similar to the slow mode component (compare Fig. \ref{fig_alfven_rho}(b) and (c)), i.e., density fluctuations propagate along the magnetic field lines. Fast modes are still present in this region, but do not make a significant contribution towards the density fluctuations. Furthermore, in the elongated region of high plasma $\beta$ along the helmet streamer axis, the propagation of density fluctuations is similar to both the slow and fast modes (compare Fig. \ref{fig_alfven_rho}(b) with panels (c) and (d)). 

The region close to the dense helmet streamer core exhibits several interesting properties because of the presence of large transverse gradients in the propagation speeds and an equipartition layer (i.e., where the sound speed equals the Alfvén speed); in terms of the plasma $\beta$, the equipartition layer is at $\beta = 1.2$ (for a monoatomic gas undergoing adiabatic expansion), but is often referred to as $\beta \approx 1$ \citep[see e.g.][]{bogdan2003waves, mclaughlin2006mhd}. The $\beta= 1$ layer is shown by the red contours in Fig. \ref{fig_alfven_rho}. In the high $\beta$ regime -- near the helmet streamer cusp -- density fluctuations propagate isotropically as both the slow and fast modes contribute significantly towards their generation (compare Fig. \ref{fig_alfven_rho}(b) with (c) and (d), within the $\beta = 1$ contour). This conversion from an anisotropic to an isotropic propagation in the high $\beta$ region occurs near the equipartition region, across which the characteristic propagation speeds of the slow and fast modes interchange -- slow waves now propagate with the Alfvén speed ($v_A$), whereas the fast waves propagate with the sound speed ($c_s$) with respect to the solar wind speed ($v_\mathrm{sw}$) \citep[][]{bogdan2003waves, mclaughlin2006mhd}. With Alfvén waves tied to the magnetic field lines and density fluctuations propagating isotropically, density fluctuations are present in the high $\beta$ current sheet of the helmet streamer (see Fig. \ref{fig_alfven_rho}(b)), where Alfvén waves cannot reach (see Fig. \ref{fig_alfven_rho}(a)).

Density fluctuations from either side of the helmet streamer interfere in the closed loop regions near the helmet streamer cusp. We note the similarity of this interference pattern with \cite{kumar2026streamer} (see e.g. their Figure 2(c)), which only considered an Alfvén wave driver at the northern footpoint of the helmet streamer. Large gradients in propagation speeds near the $\beta = 1$ layer also results in refraction and reflection of the density fluctuations. While they also contribute to the interference pattern in the current sheet as well as in the low $\beta$ region, their contributions remain secondary to those directly from the wave driver in the present simulation. 

\subsection{Wave propagation in the pseudo-streamer} \label{sec_ps}
\begin{figure*}
    \centering
    \includegraphics[width=\linewidth]{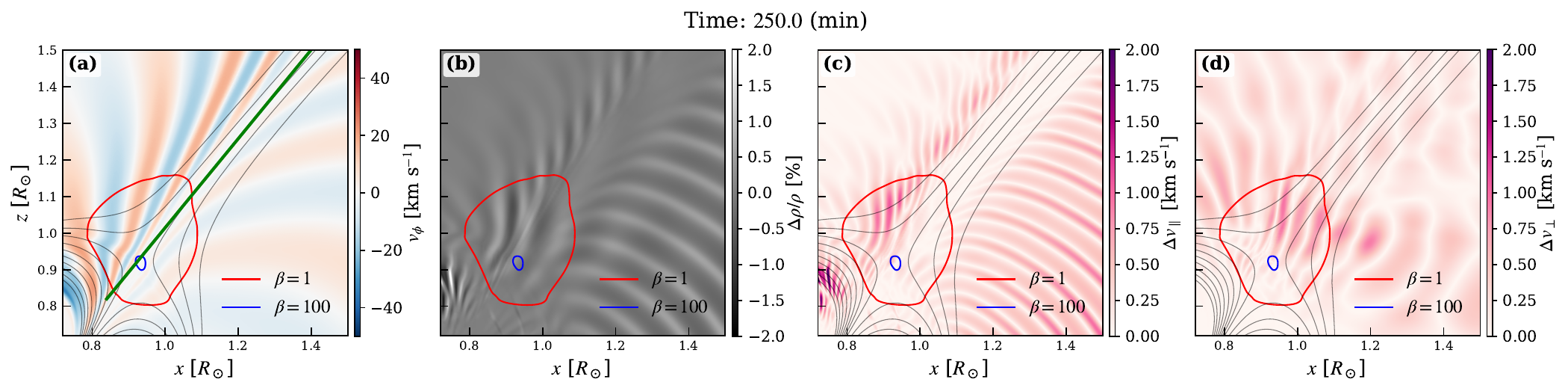}
    \caption{A zoom-in of Fig. \ref{fig_alfven_rho}, showing wave dynamics near the null point of the pseudo-streamer. The red and blue contours denote the high $\beta$ region -- $\beta = 1$ and $100$ respectively. The null point is embedded in this high $\beta$ region, where the magnetic field practically vanishes, resulting in a high plasma $\beta$.}
    \label{fig_alf_rho_roi2}
\end{figure*}

The main difference between a pseudo-streamer and a helmet streamer is the magnetic field configuration. A pseudo-streamer has the same magnetic polarity across its separatrix lying above an X-type null point, whereas a helmet streamer has opposite polarities across its axis. Alfvén waves driven at both the footpoints of the pseudo-streamer, as in the helmet streamer, undergo continuous reflections due to varying plasma characteristics along the magnetic field lines. Along the open field lines in the northern flank, the Alfvén wave amplitudes increase from $\sim 10\, \mathrm{km\, s^{-1}}$ in the low corona to $\sim 20\, \mathrm{km\, s^{-1}}$ at $\sim 1.5\, R_\odot$ as they encounter a monotonically decreasing density. Similarly, Alfvén wave amplitudes also increase along the field lines in the southern flank. However, the increase is less than that in the northern flank because of a significantly higher magnetic field expansion between the pseudo-streamer and the helmet streamer (Fig. \ref{fig_expansion_factor}), which redistributes Alfvén wave energy over a larger volume. This is also responsible for a higher Alfvén speed (and thus wavelength) in this region (Fig. \ref{fig_alfven_rho}(a)), as the density becomes smaller (Fig. \ref{fig_driver_amplitude}). Furthermore, Alfvén waves originating from opposite footpoints of the pseudo-streamer remain separated from each other since the magnetic field lines are parallel to its separatrix. On the highly curved field lines close to the separatrix, Alfvén waves encounter higher densities closer to the pseudo-streamer axis. As a result, Alfvén waves propagating along those field lines slow down, similar to the helmet streamer.

While Alfvén waves in the open field region of a pseudo-streamer exhibit qualitatively similar behaviour to those in a helmet streamer at higher altitudes, their propagation differs due to the presence of a double arcade system and a null point. Along the closed loops below the null point, Alfvén waves reflect from the other loop footpoint. This will lead to wave interference along the loops. Since the loop length is approximately equal to the Alfvén wave wavelength ($\sim 400-500\, \mathrm{Mm}$, for loops in the driver region), standing modes form in this region and an interference pattern is no longer visible, unlike in the closed loops associated with the helmet streamer. Near the null point, there is a combination of these effects -- lowering of the Alfvén speed and lack of numerical resolution -- as discussed above. Alfvén waves cannot pass through the null, since the magnetic field, which guides the waves, practically vanishes close to the null and diverges in nearby regions. (Note that although the null is defined as the point where the magnetic field becomes zero, in simulations, it is a region of extremely low magnetic field strength.) Instead, the Alfvén waves propagate away, either towards the loop footpoint, or along the open field line, eventually resulting in significant deformation of the wavefronts. As a result, there is a loss of Alfvén wave signal near the null point. 

Once again, density fluctuations are generated via Alfvén wave mode conversion to slow and fast magnetoacoustic modes. However, a comparison of Fig. \ref{fig_alfven_rho}(b) with panels (c) and (d) suggests that the slow modes have a higher contribution to the density fluctuations in the majority of the domain where the plasma $\beta$ is low. Consequently, the density fluctuations due to the slow waves propagate anisotropically along the magnetic field lines in this region. Since the slow waves are largely guided along the magnetic field lines (similar to the Alfvén waves), density fluctuations due to them remain separated across the pseudo-streamer axis. With the Alfvén waves being driven at both footpoints of the pseudo-streamer, slow waves due to them can only come close to the pseudo-streamer axis but cannot cross it. This is shown in Fig. \ref{fig_alf_rho_roi2}(c), where we have zoomed-in on a region close to the pseudo-streamer. On the other hand, density fluctuations due to the fast waves can propagate across the magnetic field lines (and also the pseudo-streamer axis, Fig. \ref{fig_alf_rho_roi2}(d)). However, as discussed in \cite{kumar2026streamer}, the contribution to the density fluctuations from fast waves in low $\beta$ regions is significantly smaller than the slow wave contribution. As a result, the amplitude of the density fluctuations along the pseudo-streamer axis is significantly lower than in neighbouring regions (Fig. \ref{fig_alf_rho_roi2}(b)), as they are only due to the fast waves. Therefore, although the fast waves are able to cross the pseudo-streamer axis, which could lead to interference, a key difference from the interference pattern observed along the helmet streamer axis is that this region of the pseudo-streamer has a low plasma $\beta$, implying that the amplitudes of such crossing signals are significantly lower.

The null point associated with the pseudo-streamer is surrounded by a localised region of high plasma $\beta$, unlike the helmet streamer, which has an elongated region of high plasma $\beta$. In this region, density fluctuations due to both fast and slow waves propagate isotropically. This is shown in Fig. \ref{fig_alf_rho_roi2}; the region within the red contour has a plasma $\beta$ greater than $1$. A comparison between Fig. \ref{fig_alf_rho_roi2}(b) and panels (c) and (d), indicates some correlation between the density fluctuations and the slow and fast wave components. Closer to the null point, the fast wave contribution to the density fluctuations increases. As such, density fluctuations from either side of the high $\beta$ region should be able to propagate across the null and generate an interference pattern similar to the one along the helmet streamer axis. However, we do not see signatures of such interference near the null point in our simulation, possibly because the amplitudes of the fast waves are too low, or the wavelengths are too small to be resolved. 

\subsection{Waves along different slits} \label{sec_slits}
\begin{figure*}
    \centering
    \includegraphics[width=\linewidth]{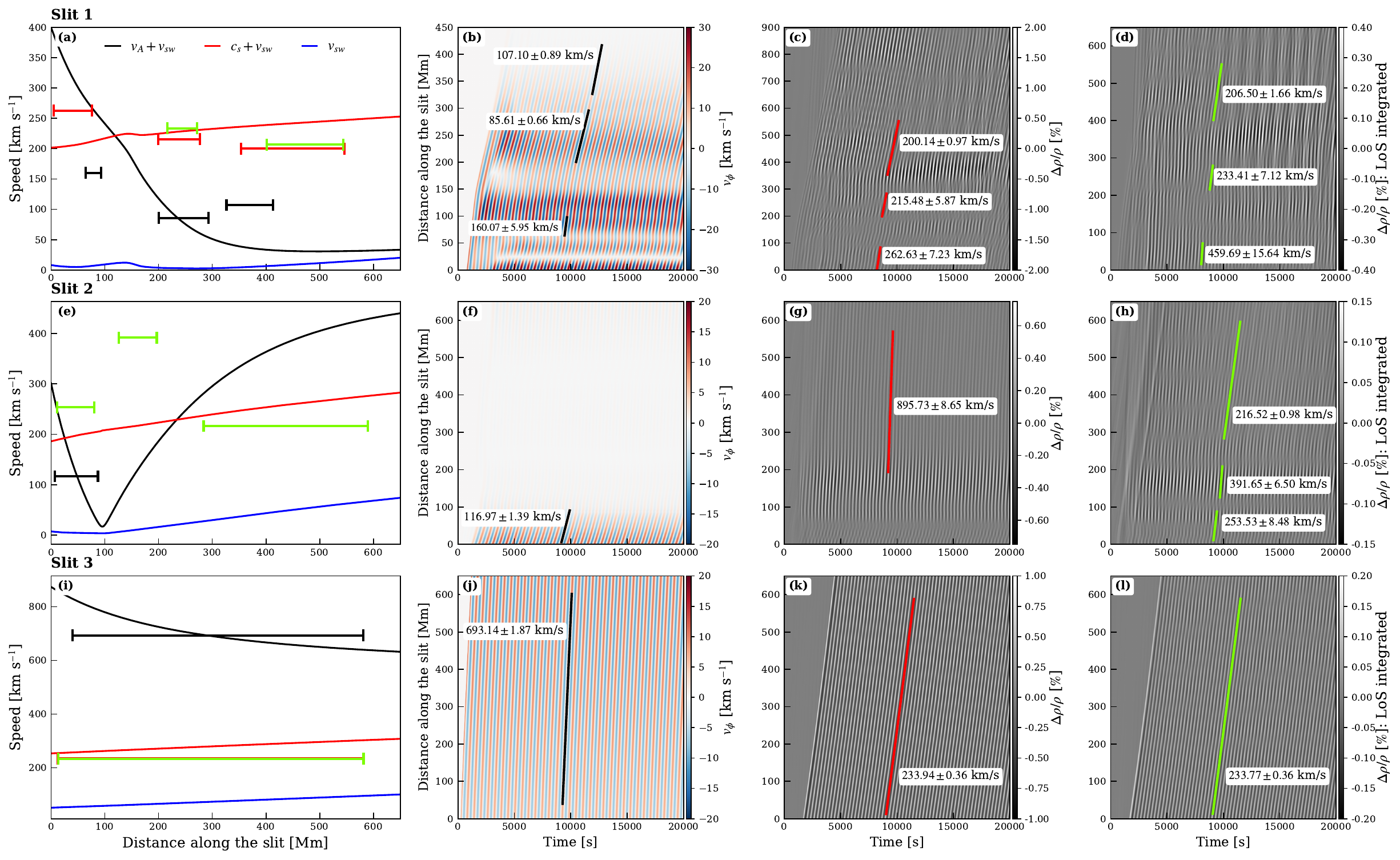}
    \caption{Time-distance maps of the Alfvén waves ($v_\phi$, second column), density fluctuations ($\Delta \rho/\rho$, third column) and the line-of-sight integrated density fluctuations (fourth column) for the three slits shown in Fig. \ref{fig_alfven_rho}(a); the time-distance maps for each slit are arrange in different rows. The first column denotes the characteristic speeds along the corresponding slit. The solid lines in the time-distance maps are linear fits, used to estimate the apparent propagation speeds of these features along these slits. The horizontal lines in the first column denote the measured apparent speeds from the time-distance maps in the adjoining columns.}
    \label{fig_td_slits}
\end{figure*}

Having discussed the qualitative similarities and differences in the propagation of the Alfvén waves and density fluctuations in the different field topologies, we will now focus on how they are captured when data along different slits are analysed, as in Metis observations. In order to compare directly with the Metis observations reported by \cite{andretta2025first}, we analyse time-distance plots for three slits chosen in the simulation domain, shown in Fig. \ref{fig_alfven_rho}(a) and numbered from $1$ to $3$. These slits are positioned to sample the domain along the helmet streamer axis (slit 1), along the pseudo-streamer passing through the null point (slit 2), and along magnetic field lines in the region between the pseudo-streamer and helmet streamer (slit 3). Slits 1 and 2 are similar to locations analysed by \cite{andretta2025first}. We do note, however, that the slits used in \cite{andretta2025first}, could be placed only approximately along the helmet streamer or pseudo-streamer axes as the helmet streamer and the pseudo-streamer are three-dimensional structures, which may not perfectly align with the plane of the sky. Further, the choice of slit $3$ is to provide observational estimates in the region between the helmet streamer and the pseudo-streamer. However, density fluctuations may not be detected by the current observational capabilities, as the open field region is relatively faint.

We use these slits as probe locations to obtain time-distance maps of the Alfvén wave component ($v_\phi$) and density fluctuations ($\Delta \rho/\rho$). The time distance maps are shown in Fig. \ref{fig_td_slits}, where the second column contains the plots corresponding to $v_\phi$ and the third column contains the plots for $\Delta\rho/\rho$ for the three slits; the first column shows different characteristic speeds as a function of distance measured along the slit. The fourth column shows the fluctuations in density integrated along a line-of-sight parallel to the rotationally invariant $\phi-$direction. To do this, we rotate the two-dimensional simulation plane about the north-south axis, assuming azimuthal symmetry, to create a three-dimensional volume. As a result, there will be some differences between the line-of-sight integrated density fluctuations and the actual density fluctuations in the simulation. Since Metis white-light coronagraph observations measure plasma density and not Alfvén wave fluctuations ($v_\phi$ and $B_\phi$), the line-of-sight integrated density fluctuations from our simulations allow us to make a qualitative comparison with, for example, the perturbations in Fig. 3 of \cite{andretta2025first}. 

From the time-distance plots in Fig. \ref{fig_td_slits}, we can calculate an apparent propagation speed by measuring the slope of the diagonal bands, indicated by the solid lines. However, it is important to note that the speeds inferred from the time-distance plots are not actual propagation speeds, but rather indicative of wave arrival at those locations, which is why we refer to them as apparent propagation speeds. These speeds would only equal the actual propagation speeds if the slit is aligned with the propagation direction. Therefore, we do not necessarily expect any correspondence between the characteristic speeds in the first column of Fig. \ref{fig_td_slits} and those obtained from the time-distance plots.

The slit placed along the helmet streamer axis (slit 1) starts from the closed field region, and samples the domain all the way up to the helmet streamer cusp. Since Alfvén waves interfere in the closed field regions, we see alternating regions of enhanced wave amplitude -- between $50-150\, \mathrm{Mm}$ and $200-300\, \mathrm{Mm}$ -- in the time-distance plot shown in Fig. \ref{fig_td_slits}(b). Between $150-200\, \mathrm{Mm}$, there is a reduced wave amplitude, due to destructive interference of Alfvén waves. As these features persist at all times, they suggest the formation of standing wave modes at those locations.  We note that similar features were seen in \cite{kumar2026streamer}, but including a second driver at the southern flank of the helmet streamer in this paper enhances wave interference. Beyond $\sim 400\, \mathrm{Mm}$, there is a loss of Alfvén wave signal as the Alfvén speed decreases (Fig. \ref{fig_td_slits}(a)). The apparent propagation speeds in different parts of the slit is due to wave arrival at those locations. Waves arrive quicker in the lower part of the slit (up to $\sim 100\, \mathrm{Mm}$) than higher up because the field lines (and therefore the distance travelled by the Alfvén waves) become longer for larger distances along the slit. 

The time-distance plot of the density fluctuation for this slit yields a higher propagation speed (Fig. \ref{fig_td_slits}(c)) in comparison with the Alfvén waves. Density fluctuations are generated through mode conversion of the Alfvén waves, as they propagate along magnetic field lines. Therefore, the propagation of the density fluctuations is not constrained to the magnetic field lines -- they can also propagate across the field lines with the fast modes -- meaning that there are contributions from plasma motions along (slow mode) and across (fast mode) the field lines. For this slit, the $\beta = 1$ layer is at $\sim 100\, \mathrm{Mm}$ measured along the slit, with a low $\beta$ region lying below this height. In the closed field region in the low $\beta$ regime ($<100\, \mathrm{Mm}$ along the slit), fast modes propagate along the slit and slow modes arrive almost perpendicular to it. Despite having a lower contribution in a low $\beta$ regime, the fast modes still contribute to the density fluctuations. Although the contribution from the fast waves would result in the propagation speed being closer to the fast speed (equal to the Alfvén speed in a low $\beta$ plasma), a significant contribution also comes from the obliquely arriving slow modes, thus reducing the overall apparent propagation speed. As a result, there is a higher apparent propagation speed of the density fluctuations than the Alfvén waves (Fig. \ref{fig_td_slits}(b)) in this part of the domain. Above the $\beta = 1$ layer ($\sim 100\, \mathrm{Mm}$) along the slit, there is a higher contribution from the fast mode to the density fluctuations. As the fast waves propagate across the magnetic field lines and along the slit in this case, this results in the apparent speed being closer to the sound speed with respect to the solar wind speed (since in a high $\beta$ regime, fast modes propagate with the sound speed).

The propagation speeds inferred from the line-of-sight integrated density fluctuations for slit $1$ exhibit a similar trend -- higher speeds in the lower parts of the slit and lower further up (Fig. \ref{fig_td_slits}(d)). As noted before, we do not expect these apparent speeds to match exactly with the speeds obtained directly from the simulation data (i.e., speeds in Fig. \ref{fig_td_slits}(b)). As this calculation involves projecting the plasma from the plane of the sky to the line-of-sight, there will be contributions from different heights to the estimated intensity. These contributions are higher for lines of sight closer to the centre than those at the limb. Despite such effects, the trend in the apparent speed remains the same, and are also consistent with those reported in \cite{andretta2025first}.

For a slit along the pseudo-streamer axis (slit 2), there is a significant loss of Alfvén wave signal beyond $\sim 100\, \mathrm{Mm}$ along the slit (Fig. \ref{fig_td_slits}(f)), which is the approximate location of the null point. In this region, the Alfvén wave amplitudes are as low as $\sim 0.1\,\mathrm{km\, s^{-1}}$ as the Alfvén waves from opposite footpoints of the pseudo-streamer do not reach the pseudo-streamer axis (since they cannot cross the field lines). The amplitudes are not exactly zero as the slit is not perfectly aligned with the (slightly curved) pseudo-streamer axis. Below the null point, on the other hand, Alfvén waves propagate towards the lower boundary, while being guided along the closed loops. As this slit is approximately aligned with the magnetic field lines, the apparent propagation speed is somewhat in agreement with the Alfvén speed along slit $2$ (Fig. \ref{fig_td_slits}(e)). 

Density fluctuations along the pseudo-streamer axis (slit 2) have an apparent propagation speed of $\sim 900\, \mathrm{km\, s^{-1}}$ (Fig. \ref{fig_td_slits}(g)), significantly higher than the typical wave speeds along this slit (Fig. \ref{fig_td_slits}(e)). This is because density fluctuations are not produced along this slit as there is no Alfvén wave signature. Instead, they arrive from neighbouring regions, with the only contribution from the fast modes, as discussed in the previous section. Slow waves, restricted to propagating along the magnetic field lines, cannot reach the pseudo-streamer axis (Fig. \ref{fig_alf_rho_roi2}(c)), as they are produced along the footpoints. As such, the time-distance plot of the density fluctuations along this slit depicts their arrival at those locations. Furthermore, with the fast mode wavefronts propagating perpendicular to this slit, the associated density fluctuations arrive virtually simultaneously, leading to artificially high apparent propagation speeds along the slit. Since this slit largely lies in a low $\beta$ region -- apart from a localised region enclosing the null point between $100-200\, \mathrm{Mm}$ -- the amplitudes of the density fluctuations remain low. Close to the null, in the high $\beta$ region, there are significant contributions from both slow and fast waves (Fig. \ref{fig_alf_rho_roi2}), resulting in an overall increase in density fluctuation amplitudes (between $100-200\, \mathrm{Mm}$ in Fig. \ref{fig_td_slits}(g)). We further note that the apparent speed calculated in Fig. \ref{fig_td_slits}(g) is roughly twice the highest speed across all observation intervals reported in \cite{andretta2025first}. However, the line-of-sight integration brings down the apparent propagation speed by nearly a factor of three (Fig. \ref{fig_td_slits}(h)), and is now in agreement with the range of values provided in \cite{andretta2025first}. 

This difference in apparent propagation speeds arises from a combination of axial rotation and the fast waves being surrounded by higher amplitude slow waves near the pseudo-streamer axis. For slit 2, axial rotation results in the slow waves being present both in the foreground and background and due to their higher amplitude in a low $\beta$ plasma, these slow waves contribute more towards the line-of-sight integrated signal than the fast waves. As a result, the apparent propagation speed calculated from the line-of-sight integrated density for slit 2 reduces significantly compared to the actual density fluctuations, as the line-of-sight integrated speed is dominated by the slow waves. This effect is less pronounced for slit 1, which mostly lies in a high $\beta$ regime; here, both slow and fast waves have approximately the same contribution to the density fluctuations.

Finally, we consider a slit a that lies between the pseudo-streamer and the helmet streamer, parallel to the magnetic field lines (slit 3, Fig. \ref{fig_alfven_rho}(a)). Because of the positioning of this slit, there is uniform Alfvén wave power throughout the time-distance map shown in Fig. \ref{fig_td_slits}(k), and the estimated apparent speed is in agreement with the Alfvén speed along this slit (Fig. \ref{fig_td_slits}(i)). As this slit entirely lies in the low $\beta$ regime, slow waves form the dominant contribution to the density fluctuations, which is also confirmed by the agreement between the apparent speed -- of both the density fluctuations (Fig. \ref{fig_td_slits}(k)) and the line-of-sight integrated density fluctuations (Fig. \ref{fig_td_slits}(l)) -- and the sound speed.

\section{Discussion and conclusions} \label{sec_discussion}
\begin{figure*}
    \centering
    \includegraphics[width=\linewidth]{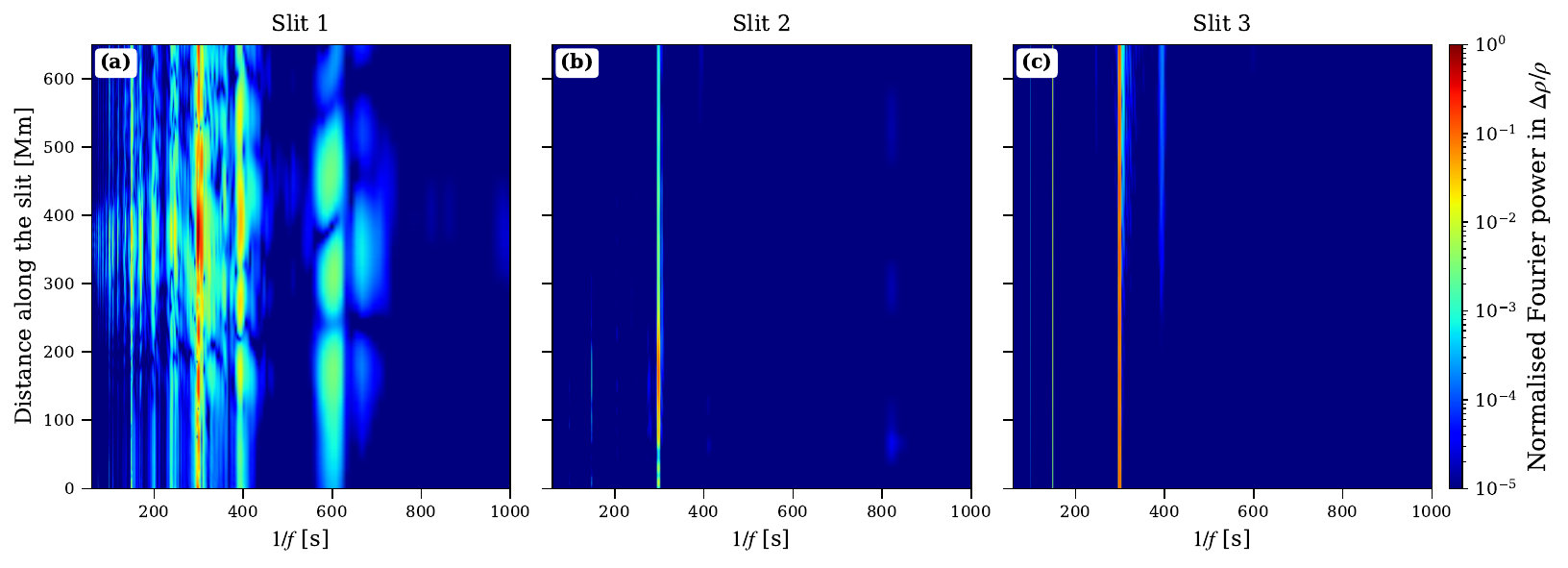}
    \caption{Fourier spectra of the density fluctuations for the three slits considered in Fig. \ref{fig_alfven_rho}(a). The spectra are normalised by the maximum power across the three panels in order to compare relative Fourier powers across periods.}
    \label{fig_slits_fft}
\end{figure*}
In this paper, we have investigated the propagation of density fluctuations produced from Alfvén waves in two distinct magnetic field configurations -- a helmet streamer and a pseudo-streamer. Our model includes key physics such as spherical expansion, gravitational stratification, thermal conduction and radiative cooling, with a supersonic and super-Alfvénic background solar wind flow. Our study is an extension of the work presented in \cite{kumar2026streamer}, which only considered a helmet streamer. Alfvén waves with a period of $10$ minutes are simultaneously driven in three distinct regions of the domain -- above the pseudo-streamer, between the pseudo-streamer and the helmet streamer and south of the helmet streamer. This extension of the study allows us to compare wave propagation between a helmet streamer and a pseudo-streamer using the same simulation, while also comparing these results with the white light observations from Metis \citep[][]{andretta2025first}.

(Non-linear) Alfvén waves can produce density fluctuations via mode conversion into slow and fast magnetoacoustic modes. This can happen due to ponderomotive effects, (non-linear) phase mixing and the presence of inhomogeneities in the medium \citep[e.g.][]{nakariakov1997alfven, verwichte1999evolution, thurgood2013ponderomotive, davies2025alfven}. However, the contribution of slow and fast wave modes to the density fluctuations depends on the plasma $\beta$; in a low $\beta$ regime, the slow waves have a higher contribution, whereas in a high $\beta$ regime, both slow and fast modes have significant contributions \citep[see][Appendix A]{kumar2026streamer}. 

A helmet streamer and a pseudo-streamer differ in their magnetic field configurations, which also results in differences in the plasma configuration. In a helmet streamer, there is an elongated high $\beta$ region along its axis, whereas a pseudo-streamer only has a localised high $\beta$ region enclosing a null point. As we illustrate, the difference in magnetic field configuration and the plasma $\beta$ profile both affect the propagation of the Alfvén waves and the induced density perturbations.

For the helmet streamer, Alfvén waves are unable to reach its axis where the current sheet lies. Despite the fact that the guiding magnetic field passes through the axis, Alfvén waves encounter a sharp decrease in the magnetic field strength towards the current sheet, which leads to a rapid decay in its amplitude. In contrast, for a pseudo-streamer, its axis coincides with the separatrix. Since Alfvén waves remain confined to the magnetic field lines, Alfvén waves are physically not able to cross the pseudo-streamer axis. However, due to the same magnetic polarity across the pseudo-streamer axis, Alfvén waves driven from opposite footpoints are able to come arbitrarily close to each other. Therefore, cross-field gradients become large, similar to phase mixing. These gradients are further increased by the asymmetry in plasma configurations across the pseudo-streamer axis. Hence, Alfvén waves driven in phase at both footpoints will not arrive simultaneously at a given height along the field lines, leading to a phase difference between Alfvén waves across the pseudo-streamer axis.

Unlike Alfvén waves, density fluctuations can propagate across magnetic field lines. Therefore, density fluctuations can be present in regions where Alfvén waves cannot reach, e.g., in the current sheet of the helmet streamer (compare Fig. \ref{fig_td_slits}(b) with (c)) and near the null point in the pseudo-streamer (compare Fig. \ref{fig_td_slits}(f) with (g)). Notably, observations from Metis have revealed that density fluctuations are ubiquitously present in the vicinity of both helmet streamers and pseudo-streamers, as in our simulation. Furthermore, counter-propagating density fluctuations from opposite footpoints of the helmet streamer and the pseudo-streamer are expected to interfere with each other.

We notice that the majority of the helmet streamer axis lies in a high plasma $\beta$ region, which allows significant propagation of the density fluctuations across the magnetic field. In contrast, the pseudo-streamer axis is embedded in a low plasma $\beta$ region, where the slow waves are absent along the pseudo-streamer axis as they cannot travel across the field. By extending the modelling to a more complex configuration where we have both a streamer and a pseudo-streamer, we learn that the density fluctuations along the pseudo-streamer axis are associated only with the fast waves, which are relatively weak in a low-$\beta$ plasma. Therefore, in our simulations, the interference near the pseudo-streamer axis is almost undetectable as it is due to the low-amplitude fast waves only. This difference -- the absence of an observable interference pattern in a pseudo-streamer -- was not seen in the Metis observations reported by \cite{andretta2025first} who found interference patterns even along the pseudo-streamer axis. Such interference patterns can potentially arise due to various effects. For example, in reality, the pseudo-streamer axis is likely to be a more complex, twisted, three-dimensional structure, which can fundamentally alter the propagation of Alfvén waves and associated density fluctuations. In such a case, density fluctuations may appear to propagate opposite to each other in overlying layers of the atmosphere, which could possibly lead to an observed interference pattern. Moreover, since density fluctuations can also propagate across magnetic field lines, there are additional possibilities for (observed) interference in a three-dimensional scenario. For density fluctuations propagating across multiple twisted magnetic field lines, we expect an enhancement of the interference pattern after a line-of-sight integration. Our two-dimensional, axially symmetric simulation and simplistic line-of-sight integration method are unable to capture these more complex dynamics.

\begin{figure*}
    \centering
    \includegraphics[width = \linewidth]{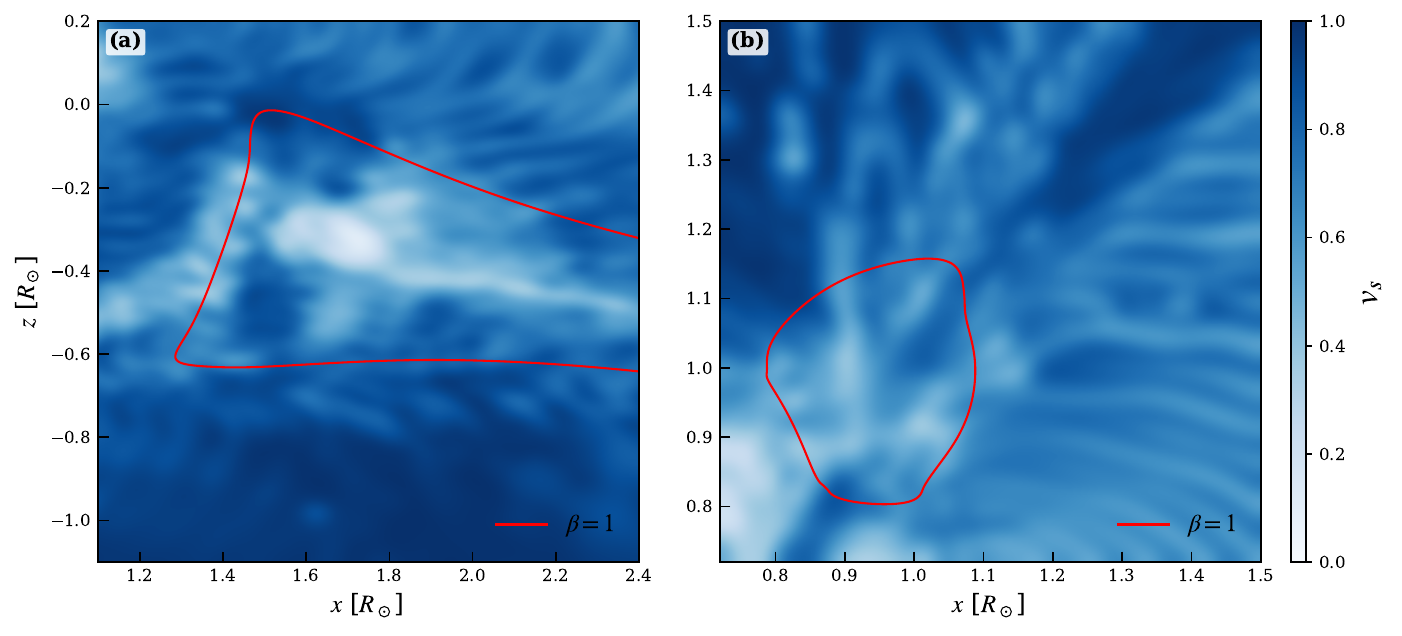}
    \caption{Contribution of the slow waves to the density fluctuations in smaller regions of the simulation domain -- (a) near the helmet streamer, and (b) near the pseudo-streamer. The solid red lines denote the $\beta = 1$ layer.}
    \label{fig_vs}
\end{figure*}

The generation of an interference pattern along the helmet streamer axis generates additional periodicities. This is illustrated in Fig. \ref{fig_slits_fft}, where we have provided Fourier powers in the density fluctuations at different heights along the slits shown in Fig. \ref{fig_alfven_rho}(a); here, the powers are normalised by the maximum power among the three panels in order to compare relative powers for different slits. Wave interference and a significant contribution from the fast modes along slit 1 (along the helmet streamer axis) redistribute the power into other frequencies (Fig. \ref{fig_slits_fft}(a)). The maximum power, however, is at $300\, \mathrm{s}$, which is half the driven Alfvén wave period ($600\, \mathrm{s}$). This indicates that the density fluctuations are predominantly generated by the ponderomotive force. On the other hand, for slit 2 (along the pseudo-streamer axis), there is only a peak at $300\, \mathrm{s}$ without significant power in other periods (Fig. \ref{fig_slits_fft}(b)). The presence of significant power at multiple frequencies in slit 1 indicates complex wave dynamics, likely involving the superposition of waves. This can give rise to interference effects, including amplitude modulation (beating) and, where reflections occur, the formation of standing wave patterns (note the alternating regions of high and low intensities in Fig. \ref{fig_td_slits}(b)). Such effects are not present along the pseudo-streamer axis, where the density fluctuations are predominantly confined to the field lines since the plasma $\beta$ is low. Instead, the Fourier power is similar to that along slit 3 (Fig. \ref{fig_slits_fft}(c)), which is fully embedded in a low $\beta$ region. While observations have shown the presence of transonic turbulence associated with density fluctuations in the near-Sun solar wind \citep[][]{zhao2025transonic}, our simulations do not capture the signatures of turbulence; the peaks in the Fourier spectra in Fig. \ref{fig_slits_fft} suggest that the energy is focused at specific frequencies -- corresponding to the monochromatic driver used in our simulations -- rather than following a power-law distribution required for a turbulent cascade.

To allow a qualitative comparison with the Metis observations, we use line-of-sight integrated densities. Assuming azimuthal symmetry, we construct a three-dimensional volume from the two-dimensional simulation plane, by rotating it about the north-south axis (i.e., in the $\phi-$direction). We note that such an assumption may not always be appropriate, particularly when there are localised flux patches in the simulation domain which do not encompass the full azimuthal direction. However, above such regions, the magnetic field (and hence the plasma distribution) tends to become predominantly radial, so at larger heights, the assumption of azimuthal symmetry remains a reasonable one. Additionally, closed loop structures are typically outside the field of view of coronagraph observations. Therefore, the line-of-sight integration described above provides a reasonable approximation. However, since pseudo-streamers are generally more localised and geometrically more asymmetric than helmet streamers, the impact of the line-of-sight integration is likely more significant for the pseudo-streamer in this study.

The line-of-sight integrated density fluctuations are $\sim 0.4\%$ for a maximum density fluctuation of $\sim 2\%$ for an output cadence of $30\, \mathrm{s}$. Observations have shown intensity fluctuations with similar amplitudes of the order of $\sim 0.1\%$ \citep[][]{andretta2025first}. In addition, the apparent propagation speeds obtained from the line-of-sight integrated density maps from our simulation range between $\sim 200-450\, \mathrm{km\, s^{-1}}$, which are also in qualitative agreement with the observations by \cite{andretta2025first}. In reality, the impact of the line-of-sight integration will also depend on aspects such as the viewing angle, which will be studied in a future paper based on a fully 3D simulation.

In this paper and in \cite{kumar2026streamer}, we have noted that the slow wave component has a higher contribution to the density fluctuations in a low $\beta$ regime, whereas both slow and fast waves have significant contributions to the density fluctuations in a high $\beta$ regime. As a result, density fluctuations predominantly follow the magnetic field lines in a low $\beta$ region, whereas their propagation becomes isotropic in a high $\beta$ region. An exception to this is the pseudo-streamer axis where, despite being in a low-$\beta$ plasma, the density fluctuations are due to the fast waves only. Therefore, the propagation of density fluctuations -- with respect to the magnetic field lines -- may be an important indicator of whether they originate from the slow or fast magnetoacoustic wave modes. From an observational perspective, this requires the knowledge of the underlying magnetic field configuration as well as the velocity fluctuations, which could be obtained either from in situ measurements or Doppler velocity measurements in conjunction with coronagraphic observations. Such observations are likely to provide the propagation speeds and wave polarisation properties that may be helpful in distinguishing density fluctuations due to slow and fast magnetoacoustic waves.

From our simulations, we can improve our understanding of the density perturbations by considering the component of the velocity which is parallel or perpendicular to the magnetic field and their correspondence to the slow and fast waves. Here, we use a heuristic approach to quantify the contributions from the parallel and perpendicular velocities to the density fluctuations. This is based on comparing their time series at each point in the simulation domain. We construct the following two quantities:
\begin{gather}
    \Tilde{v}_s \equiv \frac{\int \left|\left( \Delta \rho/\rho \right) \right| \Delta v_\parallel\, \mathrm{d}t}{\int \left(\Delta v_\parallel\right)^2\, \mathrm{d}t},\\
    \Tilde{v}_f \equiv \frac{\int \left|\left( \Delta \rho/\rho \right) \right| \Delta v_\perp\, \mathrm{d}t}{\int \left(\Delta v_\perp\right)^2\, \mathrm{d}t}.
\end{gather}
Here, $\Delta \rho/\rho$, $\Delta v_\parallel$ and $\Delta v_\perp$ denote the time series of the density fluctuations, parallel and perpendicular velocities, respectively, whose means have been subtracted from the original signal (this is done in order to only compare fluctuations). Further, we normalise $\Tilde{v}_s$ and $\Tilde{v}_f$ to calculate relative contributions:
\begin{gather}
    v_s = \frac{\Tilde{v}_s}{\Tilde{v}_s + \Tilde{v}_f}, \label{eq_vs}\\
    v_f = \frac{\Tilde{v}_f}{\Tilde{v}_s + \Tilde{v}_f},
\end{gather}
such that $v_s + v_f = 1$. The plots of $v_s$ in two smaller regions of the simulation domain -- close to the helmet streamer and close to the pseudo-streamer -- are provided in Fig. \ref{fig_vs}; here, a 2D Gaussian filter is applied to smooth out fluctuations (on the scale of cell size) in $v_s$. In general, in Fig. \ref{fig_vs}, we note a lower slow wave contribution in the high $\beta$ region (marked by the $\beta = 1$ contour in red) -- $v_s < 60\%$ -- meaning that the perpendicular velocity contribution increases to more than $40\%$ in those regions. Therefore, this high $\beta$ region has approximately equal contributions from both parallel and perpendicular velocity components. In contrast, regions in the low $\beta$ plasma have a higher contribution from the parallel component, with $v_s$ going as high as $\sim 80\%$.

We further note a lower parallel velocity contribution in the low $\beta$ region near the $\beta = 1$ layer in Fig. \ref{fig_vs}. Although we do not aim to disentangle the multiple effects at play that generate density fluctuations in our simulation, this lower contribution from the parallel velocity is expected to be due to the presence of large transverse gradients near the closed loops of the helmet streamer and the pseudo-streamer, which can generate perpendicular velocities through non-linear phase mixing of the Alfvén waves and thus contribute to $v_f$. Overall, we find a good correlation between parallel and perpendicular velocity contributions and the plasma $\beta$ in the simulation domain. Given the correlation we find, these results raise the possibility of using parallel and perpendicular decomposition of the velocity field as a proxy to estimate the plasma $\beta$ of the background plasma.

The work presented in this paper compares the propagation of Alfvén waves and associated density fluctuations in a helmet streamer and a pseudo-streamer. The differences in their propagation arise due to the different magnetic field configurations, but these differences seem to blur when a line-of-sight density integration is performed across different slits in order to compare them with white light Metis observations \citep[][]{andretta2025first}. In general though, the generation of density fluctuations from Alfvén waves are due to the same set of mode conversion mechanisms -- the ponderomotive effect and the presence of transverse gradients and inhomogeneities. Although complete disentanglement of the various wave modes is not straightforward in a complex plasma configuration such as ours, we have been able to provide estimates of the relative contributions from the slow and fast mode waves. The limitations associated with the 2D nature of our simulations will be addressed in a forthcoming fully 3D study.

\section*{Acknowledgements}
AK and SD-Y received funding from the Leverhulme Trust (RPG-2021-114). TAH received funding from the Carnegie Trust grant number RIG013386. PP acknowledges support by the Italian PRIN 2022, project 2022294WNB entitled ``Heliospheric shocks and space weather: from multispacecraft observations to numerical modeling”, funded by Next Generation EU, within Piano Nazionale di Ripresa e Resilienza (PNRR), Missione 4 ``Istruzione e Ricerca” - Componente C2 Investimento 1.1, `Fondo per il Programma Nazionale di Ricerca e Progetti di Rilevante Interesse Nazionale (PRIN)'. IDM received funding from the Research Council of Norway through its Centres of Excellence scheme, project number 262622. VA acknowledges support by Addendum n. 1 to the ASI-INAF agreement n. 2018-30-HH.0 for the science support to Metis, SWA DPU e STIX in the D/E phases. The authors also thank the anonymous referee for their valuable comments.

\section*{Data availability}
The data presented in this article will be available from the University of St Andrews research repository.

\bibliographystyle{mnras}
\bibliography{bibliography}

\bsp	
\label{lastpage}

\end{document}